\documentclass[runningheads]{llncs}
\usepackage[T1]{fontenc}
\usepackage{graphicx}
\usepackage{booktabs}
\usepackage{tabularx}
\usepackage{array}
\newcolumntype{P}[1]{>{\centering\arraybackslash}p{#1}}

\newcommand{\opus}[1]{%
  \begingroup
    \spaceskip=\fontdimen2\font plus \fontdimen3\font minus \fontdimen4\font
    \xspaceskip=\fontdimen7\font\relax
    \ttfamily
    #1%
  \endgroup
}

\begin{document}
\title{When Bias Backfires:
\newline The Modulatory Role of Counterfactual Explanations on the Adoption of Algorithmic Bias in XAI-Supported Human Decision-Making}

\titlerunning{When Bias Backfires}

\author{Ulrike Kuhl\inst{1}\orcidID{0000-0002-9405-918X} \and
Annika Bush\inst{2}\orcidID{0000-0003-1475-4260}}
\authorrunning{U. Kuhl and A. Bush}
\institute{Machine Learning Group, Bielefeld University, Universitätsstr. 25, 33615 Bielefeld, Germany
\email{ukuhl@techfak.uni-bielefeld.de}\\
\url{https://hammer-lab.techfak.uni-bielefeld.de}
\and
Research Center Trustworthy Data Science and Security of the University Alliance Ruhr, Faculty of Informatics, Technical University Dortmund, Joseph-von-Fraunhofer-Str. 25, 44227 Dortmund, Germany
\email{annika.bush@tu-dortmund.de}}
\maketitle              % typeset the header of the contribution
\begin{abstract}

Although the integration of artificial intelligence (AI) into everyday tasks improves efficiency and objectivity, it also risks transmitting bias to human decision-making.
In this study, we conducted a controlled experiment that simulated hiring decisions to examine how biased AI recommendations - augmented with or without counterfactual explanations - influence human judgment over time. 
Participants, acting as hiring managers, completed 60 decision trials divided into a baseline phase without AI, followed by a phase with biased (X)AI recommendations (favoring either male or female candidates), and a final post-interaction phase without AI.
Our results indicate that the participants followed the AI recommendations 70\% of the time when the qualifications of the given candidates were comparable.
Yet, only a fraction of participants detected the gender bias (8 out of 294).
Crucially, exposure to biased AI altered participants' inherent preferences: in the post-interaction phase, participants’ independent decisions aligned with the bias when no counterfactual explanations were provided before, but reversed the bias when explanations were given.
Reported trust did not differ significantly across conditions.
Confidence varied throughout the study phases after exposure to male-biased AI, indicating nuanced effects of AI bias on decision certainty.
Our findings point to the importance of calibrating XAI to avoid unintended behavioral shifts in order to safeguard equitable decision-making and prevent the adoption of algorithmic bias.
In the interest of reproducible research, study data is available at: \url{https://github.com/ukuhl/BiasBackfiresXAI2025}.

\end{abstract}
\section{Introduction}
The increasing adoption of artificial intelligence (AI) in hiring processes promises enhanced efficiency and objectivity in candidate selection. %However, this integration raises critical questions about how AI systems influence human decision-making, particularly when they are biased, and thus unduly favor or disadvantage certain individuals or groups~\cite{kordzadeh2022algorithmic}.
Yet, it also raises critical questions about the impact of biased AI - unduly favoring or disadvantaging certain individuals or groups~\cite{kordzadeh2022algorithmic} - on human judgment.

While considerable research and technical advances have focused on detecting and mitigating this so-called algorithmic bias~\cite{bellamy2019ai,huang2024contextual,pulivarthy2025bias},
%less attention has been paid to understanding how biased AI recommendations might shape human judgment over time. 
less is known about how biased AI recommendations affect human decision-making.
One seminal study demonstrates that humans may indeed internalize algorithmic bias from mere interaction with AI recommendations~\cite{vicente2023humans}.
Yet, the potentially critical role of explainability in this context remains unclear.
Designed to be user-friendly and facilitate seamless interaction with the system, explanations may inadvertently cause users to uncritically follow biased recommendations.
Within this broader concern, counterfactual explanations (CEs) are particularly noteworthy.
These explanations present actionable ``what-if'' scenarios, designed to make AI decisions particularly transparent in a way that resembles human cognition~\cite{byrne2019counterfactuals,wang2021counterfactual}. 
However, by providing alternative scenarios to justify outcomes, they may paradoxically increase the likelihood that humans adopt AI-inherent biases:
When an AI system provides seemingly logical alternative scenarios to justify its recommendations, users may be more inclined to accept these suggestions, even when they conflict with their own judgment or objectively fair decisions. 
On the other hand, their potential positive effects include enhancing the estimation of an AI model's accuracy, reducing over-reliance on erroneous outputs, and improving human-AI collaborative decision-making~\cite{lee2023understanding}.
Overall, both potential dynamics are highly relevant in hiring contexts, where decisions have lasting impacts on individuals and organizational diversity.

%Previous research has extensively documented the presence of bias in AI recruitment systems \cite{rosenthal2024michael,pulivarthy2025bias}.
%Similarly, studies have demonstrated how XAI can enhance user trust and decision-making in some contexts \cite{Biloborodova.2023}. However, the specific impact of CEs on the transmission of algorithmic bias to human decision-makers remains largely unexplored.
Prior studies have documented bias in AI recruitment systems~\cite{rosenthal2024michael,pulivarthy2025bias} and shown that XAI can enhance decision-making in certain settings~\cite{Biloborodova.2023}.
Yet, the effect of CEs on bias transmission remains unclear.
Our research addresses this gap through a controlled experiment examining how AI recommendations, both with and without CEs, influence hiring decisions. We specifically investigate whether exposure to biased AI recommendations leads participants to make more biased decisions in subsequent independent evaluations. Furthermore, we examine whether XAI affects users' likelihood to accept AI recommendations and their trust in the system. Through this investigation, we aim to contribute three key insights to the XAI community:

\begin{enumerate}
    \item Understanding of how AI bias can impact human decision-making through repeated interaction.
    \item Examining XAI's role in either facilitating or preventing bias transmission.
    \item Highlighting implications for AI systems that support human decision-making while protecting against bias adoption.
\end{enumerate}

%As organizations increasingly rely on AI to support hiring decisions, understanding how these systems influence human judgment becomes crucial for maintaining fair and equitable selection processes.
%Our results suggest specific approaches for calibrating trust in AI recommendations and designing interfaces that promote appropriate skepticism rather than blind acceptance.

%The remainder of the paper reviews related literature, describes the experimental methodology, presents our results, and discusses implications for theory and practice, with recommendations for future research and system design.

\section{Related Work}
\subsection{AI Bias in Decision Support Systems}
As AI increasingly penetrates organizational decision-making processes, understanding the manifestation and transmission of AI bias has become crucial, particularly in high-stakes domains like hiring. %Decision support systems (DSS) promise to enhance human judgment through data-driven insights, yet mounting evidence suggests these systems may inadvertently perpetuate or even amplify existing biases.
While decision support systems (DSS) are intended to improve decision-making, they may inadvertently reinforce existing biases.

%The challenge of AI bias in DSS emerges from multiple interacting sources.
AI systems can inherit biases present in their training data,
%leading to systematically discriminatory outcomes against certain demographic groups \cite{rosenthal2024michael}.
thus producing subtle, yet discriminatory outcomes~\cite{rosenthal2024michael}.
%These algorithmic biases often manifest subtly, making them particularly difficult for users to detect and correct. 
%Pulivarthy and Whig \cite{pulivarthy2025bias}
%demonstrate how seemingly neutral feature selection in hiring algorithms can disadvantage qualified candidates based on gender or ethnicity, even when protected attributes are explicitly excluded from the model.
Even neutral feature selection may disadvantage qualified candidates by gender or ethnicity without explicitly including protected attributes~\cite{pulivarthy2025bias}.
At the algorithmic level, different model choices elicit distinct model behavior, inadvertently biasing outcomes~\cite{hooker2021moving}.  
%More concerning is the emergence of automation bias, where users demonstrate excessive trust in AI recommendations despite evidence of system limitations. Recent work by Kücking et al. \cite{kucking2024automation} reveals that professionals across domains tend to defer to AI suggestions even when these conflict with their own judgment, particularly when under time pressure or cognitive load. This finding has profound implications for hiring contexts, where decision-makers must evaluate complex, multifaceted candidate profiles while managing organizational pressures.
Another serious concern arises when professionals defer to AI suggestions, even when these conflict with their judgment~\cite{kucking2024automation}. This so-called automation bias is aggravated by time pressure, posing serious risks in domains like hiring.

%The user interface itself can either mitigate or exacerbate these biases. Militello et al. \cite{Militello.2025} document how interface design choices — from the presentation of confidence scores to the ordering of recommendations — significantly influence user behavior and decision-making processes. Their work suggests that even subtle design decisions can shape how users interpret and apply AI recommendations, potentially amplifying existing biases.

Current approaches to addressing AI bias in DSS focus on multiple intervention points.
%At the technical level, researchers are developing fairness-aware algorithms that explicitly optimize for equitable outcomes across demographic groups \cite{pulivarthy2025bias}. Huang and Zaslavsky \cite{Huang.2024} propose innovative bias detection frameworks using contextual knowledge graphs, enabling real-time monitoring and mitigation of biased recommendations.
%However, technical solutions alone prove insufficient.
Efforts to mitigate AI bias include fairness-aware algorithms that explicitly optimize for equitable outcomes across demographic groups~\cite{pulivarthy2025bias}, and bias detection frameworks using contextual knowledge graphs~\cite{Huang.2024}. 
Complementary research from a socio-technical perspective emphasizes the importance of user training and education in reducing over-reliance on AI recommendations~\cite{kucking2024automation}.
%Kücking et al. \cite{kucking2024automation} demonstrate that comprehensive training programs can help users maintain appropriate levels of skepticism toward AI suggestions, particularly among those without technical backgrounds.
%This finding suggests that effective bias mitigation requires a socio-technical approach that addresses both system design and user behavior.

Despite these advances, completely eliminating AI bias in DSS remains an elusive goal.
The complex interplay between algorithmic biases and human cognition creates challenges that resist simple solutions.
This reality underscores the importance of understanding how biases propagate through human-AI interaction, particularly in contexts where decisions have lasting social impact.
\subsection{Human-AI Decision Making}
The dynamics of human-AI collaboration in professional decision-making extend beyond simple questions of accuracy or efficiency. As organizations increasingly deploy AI systems to support hiring and other critical decisions, understanding the intricate patterns of human-AI interaction becomes essential for designing effective collaborative systems.

Trust emerges as a foundational element in collaborative human-AI decision-making. When users trust AI systems appropriately, decision accuracy can improve significantly~\cite{Biloborodova.2023}. However, this trust must be calibrated carefully, requiring users to understand both the capabilities and limitations of the AI system. Explainability plays a crucial role in this calibration process, allowing users to validate AI suggestions against their own expertise and domain knowledge~\cite{Biloborodova.2023}.

The concept of complementary team performance provides a useful framework for understanding successful human-AI collaboration, as information and capability asymmetries between humans and AI systems can enhance decision quality~\cite{Hemmer.2024}. In hiring contexts, this might manifest as AI systems excelling at analyzing quantitative credentials, while humans better assess cultural fit and interpersonal dynamics. The key lies in designing interfaces and workflows that capitalize on these complementary strengths, rather than forcing either party to operate outside their optimal domain.

The development of effective human-AI partnerships depends heavily on human learning processes.
Users must develop accurate mental models of AI capabilities through experience~\cite{schemmer2023towards}. 
In line with this, Suffian et al.~\cite{suffian2023cl} developed the CL-XAI system, integrating user-feedback-based CEs that facilitate the development of accurate mental models of the AI's capability through direct human-AI interaction.
%This concept of experiential learning proves particularly crucial in professional contexts like hiring, where decision-makers must learn to appropriately weight AI recommendations against their own judgment. Their research suggests that this learning process is not purely cognitive but involves developing intuitive understanding through repeated interaction with AI systems.
Still, significant challenges remain in optimizing these collaborative relationships. The risk of over-reliance on AI recommendations—particularly when they align with existing biases or preferences—can undermine the potential benefits of human-AI collaboration~\cite{RasiklalYadav.2024}. Similarly, misinterpretation of AI outputs, especially in complex decision spaces like candidate evaluation, may lead to suboptimal outcomes even when the underlying AI system performs well.

These challenges highlight the need to understand how humans interpret and integrate AI recommendations, where trust in AI may shape judgment over time and subtly reinforce human biases through repeated interaction.

\subsection{The Role of XAI and Counterfactual Explanations}

Explainable AI (XAI) has emerged as a promising solution to address concerns about transparency and accountability.
Certain XAI methods—particularly nearest-neighbor examples—enable users to develop a more nuanced understanding of when to trust or when to question AI recommendations~\cite{Humer.2024}.
While the initial acceptance of intelligent systems is largely driven by their performance, transparency plays a critical indirect role in fostering trust, enhancing users' perceptions of reliability and competence that ultimately increase their willingness to rely on these systems~\cite{wanner2022effect}.

However, despite its promise, recent research shows that XAI's impact on user behavior is multifaceted and can sometimes yield unintended consequences.
For instance, certain approaches may inadvertently promote over-reliance or even legitimize biased outputs~\cite{RasiklalYadav.2024}.
In fact, evidence suggests that XAI may mislead users by fostering undue trust in black box systems, even though these systems are concurrently perceive as not trustworthy~\cite{lakkaraju2020fool}.

Within the broader landscape of XAI approaches, CEs have emerged as a particularly powerful mechanism with potential impacts on user trust and decision-making.
By illustrating how different inputs could lead to alternative outcomes (`If feature X had been different, the outcome would have been different.'), they provide users with actionable insights into AI system behavior.
Notably, previous user-based assessments show that CEs offer concrete decision-support enabling users to adjust their behavior to improve task performance~\cite{kuhl2022keep,kuhl2023let}.

There is evidence that CEs may play a crucial role in trust calibration between users and AI systems. Del Ser et al. \cite{DelSer.2024} demonstrate how CEs help users develop more nuanced understanding of a system's reliability by explicitly showing the conditions under which predictions might change. This understanding becomes particularly crucial in high-stakes decisions, where CEs help users develop more appropriate levels of trust in AI recommendations~\cite{lee2023understanding}.
Similarly, Rüttgers et al.~\cite{ruttgers2024automatic} demonstrate in an automated matchmaking context that CEs can significantly enhance user trust and system transparency.

However, findings regarding the positive effects of CEs on user trust remain mixed.
A number of  user-based-evaluations fail to demonstrate significant improvements in trust driven by CEs~\cite{wang2021explanations,scharowski2023exploring}.
Moreover, research has shown that CEs must be properly calibrated to align with the specific task and human preferences, both regarding the types of features used and the counterfactual's directionality~\cite{warren2024categorical,kuhl2023better}.
Papenmeier et al.~\cite{papenmeier2022s} found that the effect of explanations on trust is contingent upon model accuracy, with faithful CEs enhancing trust only in high-accuracy scenarios. 
This further highlights the complex relationship between CEs and trust, meriting careful evaluation. 

Particularly in DSS contexts, CEs may help reduce harmful over-reliance on AI.
In clinical settings, Lee and Chew \cite{Lee.2023} documented a 21\% reduction in user over-reliance when presenting CEs compared to feature importance scores. Thus, showing users alternative scenarios might encourage more critical evaluation of AI recommendations rather than blind acceptance.
Further, recent technological advances have further enhanced the potential impact of CEs in DSS contexts. Interactive systems like FACET~\cite{VanNostrand.2024} and user-feedback-based CEs~\cite{suffian2023cl} enable users to actively explore different scenarios, providing practical guidance that bridges the gap between AI outputs and human reasoning.
%This interactivity appears to strengthen users' sense of agency in the decision-making process, potentially leading to more thoughtful consideration of AI recommendations.

However, the effectiveness of CEs may have an unexpected downside. When explanations appear particularly logical or compelling, they might actually increase users' tendency to accept potentially biased recommendations or unfair systems~\cite{lakkaraju2020fool}. This risk may become especially relevant in hiring contexts, where seemingly rational explanations for biased decisions could make discriminatory patterns harder to detect and resist.
This conflict between counterfactuals' capacity to foster critical evaluation and their potential to legitimize biased recommendations highlights the need for careful investigation of their impact on human decision-making. %Understanding how users process and apply counterfactual insights becomes crucial for designing systems that genuinely support unbiased decision-making rather than inadvertently reinforcing problematic patterns.

\subsection{Hypotheses}

Based on the literature review, we formulate the following hypotheses:
\begin{description}
    \item[H1:] Participants receiving XAI CEs will show different rates of agreement with AI recommendations compared to those receiving black-box AI recommendations.
    \item[H2:]
    Interaction with biased (X)AI recommendations will shift participants' gender-based decision patterns in subsequent independent evaluations, increasing alignment with the (X)AI's bias direction.
    \item[H3:] Decision confidence changes depending on the (X)AI decision recommendations and the induced gender bias (male/female).
    \item[H4:] Participants receiving XAI CEs will show higher trust in AI recommendations compared to those receiving black-box AI recommendations.
\end{description}

\section{Method}
\subsection{Study Design}
We employ a randomized controlled experiment to test whether exposure to biased (X)AI recommendations influences human decision-making in hiring scenarios.
The experiment consists of four distinct phases designed to measure how (X)AI recommendations influence human decision-making in hiring contexts. 
In the experimental scenario, participants were asked to assume the role of a hiring manager at a large company, tasked with reviewing and evaluating pairs of candidates based on their qualifications for a new position.
Candidate information presented included the aptitude features experience, references, soft skills, and education, along with a professional headshot of the candidate.

In phase 1 (baseline assessment), participants establish their baseline decision-making patterns by reviewing 20 pairs of candidate profiles without (X)AI assistance. 
For each pair, participants are asked to make a hiring decision for one of the candidates based on the information presented.
This phase serves as a crucial baseline measure of any pre-existing biases.

\begin{figure}[t]
\includegraphics[width=\textwidth]{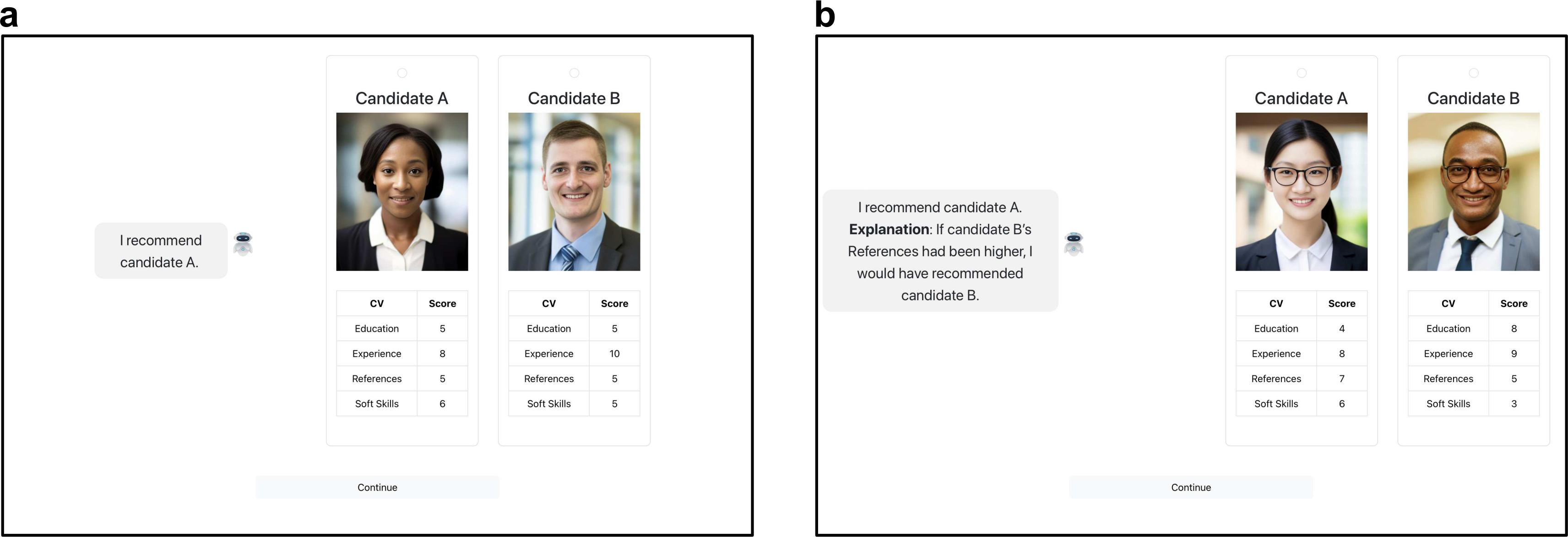}
\caption{Materials presented to participants during the (X)AI-interaction phase. Participants indicate their choice by clicking on one of the two candidate boxes. a) Exemplary decision trial shown for groups in the AI condition. b) Exemplary decision trial shown for groups in the XAI condition, featuring a counterfactual explanation.} \label{fig:AI-Interaction}
\end{figure}

Phase 2 -- (X)AI interaction -- introduces participants to an (X)AI recruitment assistant and represents the core experimental manipulation. Participants review 20 new pairs of candidates, this time with (X)AI recommendations. 
Participants in the AI conditions receive only the AI's recommendations (Fig.~\ref{fig:AI-Interaction}a), while those in the XAI conditions receive the same recommendations supplemented with a CE (Fig.~\ref{fig:AI-Interaction}b). 
In addition to the (X)AI manipulation, the recommendations participants encountered were systematically biased either against female or male candidates.
%Then both groups provide confidence ratings for their decisions. The (X)AI system is programmed to systematically favor one gender, though this bias is calibrated to remain subtle.

In phase 3 (post-interaction decisions), participants return to making decisions without (X)AI assistance, reviewing another set of 20 candidate pairs.
This phase is critical to measuring any persistent effects of (X)AI exposure on decision-making patterns.

At the end of each of the decision phases 1-3, participants indicate their confidence in their decision-making on a 5-point Likert scale ranging from ``Not at all confident'' to ``Extremely confident''.

The experiment concludes with phase 4 (post-study assessment), where participants complete a brief questionnaire about their demographics (age and gender identity) and their trust in the (X)AI assistant. Additionally, they may respond to two open-ended questions assessing a) whether they noticed anything about the (X)AI tool and b) their understanding of the purpose of the study (Table~\ref{tab:trust-scale}). 
The Ethics Committee of Bielefeld University, Germany, approved this study.

%Throughout phases 1-3, the CV pairs are carefully constructed to control for qualifications, experience, and other relevant factors, with gender serving as the primary variable of interest. The progression through these phases allows us to track changes in decision-making patterns and potential bias adoption from pre-(X)AI exposure through post-(X)AI interaction.

\subsection{Materials Development}

\subsubsection{Image Stimuli}

\begin{figure}[t]
\includegraphics[width=\textwidth]{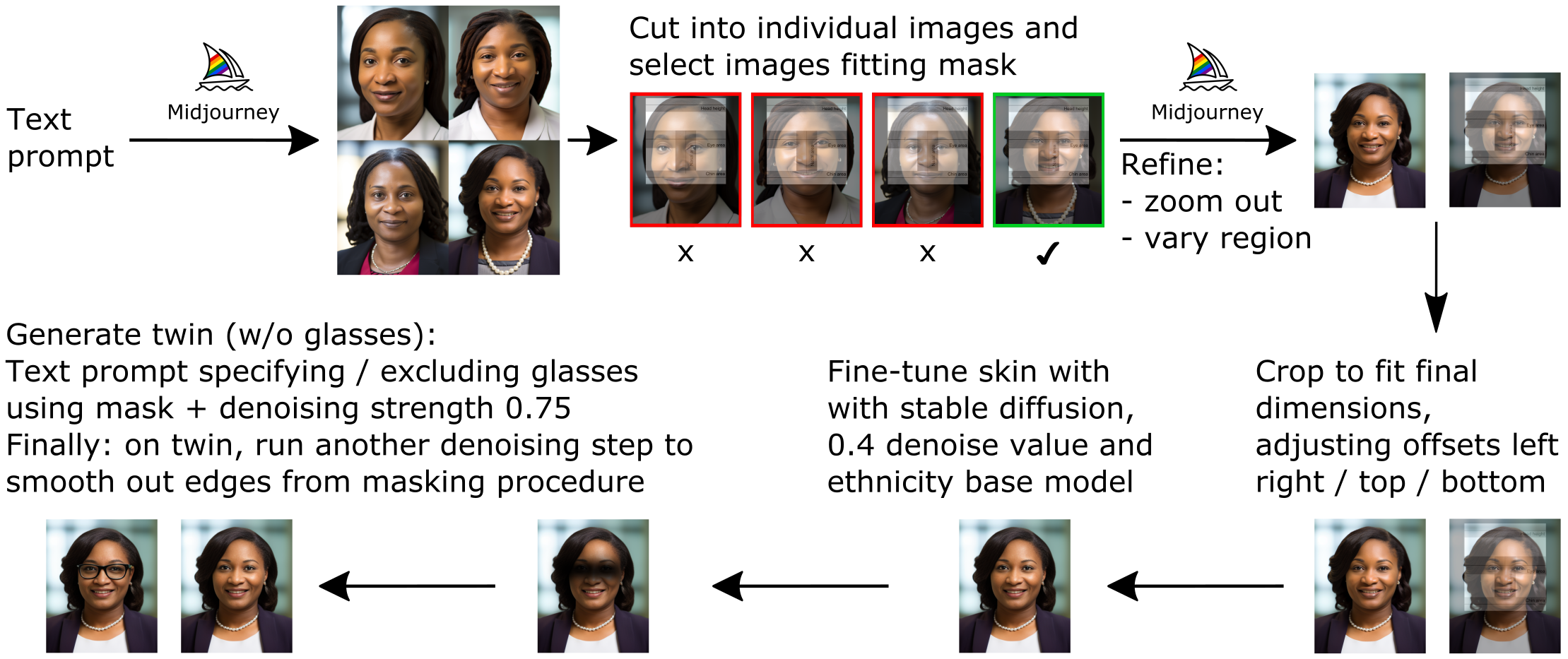}
\caption{Image generation pipeline.} \label{fig:ImageGeneration}
\end{figure}

We developed a multi‐stage pipeline to generate and process a stimulus set of professional headshot images using a combination of generative models and image processing tools (Fig.~\ref{fig:ImageGeneration}).
The pipeline comprised initial image generation with Midjourney\footnote{https://www.midjourney.com/home} and subsequent refinement using Stable Diffusion~\cite{rombach2022high} in an inpainting framework.

In the first stage, approximately 40 images per demographic group (i.e. females / males of European, African, and East Asian descent, respectively) were generated with Midjourney.
Groups were defined by ethnicity (European, Asian, African descent), gender, and the presence or absence of glasses.
The images were constrained to depict individuals in their 30s with an average appearance in a professional business headshot format, set against an indoor, blurred office background. To achieve hyperrealistic results, prompts were carefully designed and parameters adjusted. For instance, the prompt for individuals without glasses was structured as follows:\\

\noindent \opus{[European | East Asian | African] [male | female] in [his | her] thirties, experienced business professional, average looking, friendly, profes- sional headshot showing head and shoulders, hyperrealistic, indoors, blurred out office setting background, portrait photography --v 5.2 --style raw --ar 5:6 --stylize 5 --c 50 --weird 50 --no glasses, pretty, good skin, beautiful, model, soft}\\

Key Midjourney parameters included \texttt{--stylize} (enforcing a closer match to the prompt),
\texttt{--chaos} or \texttt{-c} (introducing variability), and
\texttt{--weird} (injecting offbeat qualities relative to prior outputs).
Exclusion lists via the \texttt{--no} parameter were used to minimize undesired artifacts, although some deviations (e.g., inadvertent inclusion of glasses) necessitated manual sorting and prompt re-submission.
Selected images were upscaled to maintain a consistent level of professionalism.
Following generation, images underwent systematic pre-processing to standardize their dimensions and align features with a predefined mask.

To further diversify the stimulus set, a final modification step was implemented using Stable Diffusion with the Automatic1111 interface for inpainting based on a pretrained model for realistic facial imagery\footnote{Realistic Vision V6.0 B1 inpainting (VAE):\\ \url{https://civitai.com/models/4201?modelVersionId=245627}}.
This step specifically targeted the addition or removal of glasses, thereby effectively doubling the dataset.
The inpainting process was configured to utilize CodeFormer for face restoration (with weight set to 0 for maximal inpainting effect), and prompt conditioning was applied as follows: a positive prompt such as \opus{eyes of an east-Asian male, 30–40 years old, rimmed glasses, experienced business professional} and a negative prompt listing undesired attributes (e.g., \opus{pretty, good skin, beautiful, model, soft}).
Batch processing was conducted with images resized to 480×576 pixels using the “resize and fill” mode and generated with the DPM++2M Karras sampling method. A final cropping step ensured that all output images conformed to the required dimensions.

This integrated pipeline -- combining controlled prompt-based generation, systematic pre-processing, and targeted inpainting -- yielded a diverse and standardized dataset of professional business headshots suitable study materials.

\subsubsection{Candidate Generation and Matching}\label{subsec:CandMatch}

We generated simulated aptitude data to create a pool of potential candidate profiles, each to be accompanied by one of the image stimuli and employment-relevant information.
Each aptitude feature (experience, education, references, and soft skills) was scored using a point system (ranging from 2 to 10), with higher scores indicating better qualifications.
In addition, each profile contained information on protected attributes: ethnicity (East-Asian, European, African) and gender (male, female).
A total of 120 candidate profiles were generated (60 per gender, consisting of 20 profiles per ethnicity, respectively). Feature scores were drawn from a normal distribution ($M$ = 7, $SD$ = 2), rounded to the nearest integer and bound between 2 and 10 to ensure realistic and moderate values.

Candidate pairs were carefully matched.
In each trial, participants viewed the profiles and images of two previously unseen candidates. In 14 out of 20 pairings per phase, candidates of opposite genders (male and female) were compared.
The remaining six pairings served as control comparisons (three male-male and three female-female), with ethnic diversity maintained across comparisons.
For biased conditions (favoring either males or females), candidate pairs were selected under two criteria: (1) the difference in total feature scores between the paired candidates was to be kept small (mean difference = -0.072, SD = 5.010 in final study), and (2) the advantaged candidate possessed at least one feature where they outperformed the marginalized candidate.
This matching criterion enabled the formulation of plausible upward CEs focused on the non-recommended candidate (e.g., ``If candidate B's Education had been higher, I would have recommended candidate B.'')~\cite{kuhl2023better}.

Finally, candidate photographs were randomly assigned to each set of aptitude scores, ensuring that the image conveyed the appropriate gender and ethnicity.
Notably, gender information was only presented through the images, as the tabular aptitude data did not include explicit gender markers.

\subsubsection{Biased Recommendations}

Participants were informed that candidate selection decisions were made by an AI assistant.
In reality, no predictive model was fitted and instead, decisions were determined by the following straightforward algorithm.
For control trials involving candidates of the same gender, the algorithm recommended the candidate with the higher total score across individual aptitude features.
If the total aptitude scores were identical, the recommendation was randomly determined.
For pairs of interest (i.e., candidates of different genders), the AI recommendation consistently favored the candidate aligned with the bias condition.
In the XAI condition, explanations were generated online by comparing the aptitude feature values for each candidate pair, identifying the differing features, and selecting one feature where the disadvantaged candidate scored lower.
The explanation conveyed that if the disadvantaged candidate's score of that feature were higher, they would have been selected instead.
It is important to note that this system is not intended to represent a fully autonomous AI solution.
Instead, we deliberately employed a controlled approach to precisely manage the recommendations and to isolate the impact of CEs that might be generated automatically by state-of-the-art AI tools.

\subsubsection{Trust Scale}
The trust scale used in our study is adapted from Hoffman et al. \cite{hoffman2023measures} and comprises 10 items measured on a 5‐point Likert scale ranging from ``I disagree strongly'' to ``I agree strongly''.
To mitigate response bias, we reversed the valence of items 2, 5, and 10 so that participants could not simply provide uniform ``agree'' responses without contradiction.
As a further quality measure, we included a repeated item with reversed wording to detect contradictory responses (item 5 and item 9 in Table \ref{tab:trust-scale}).
Additionally, an attention check was incorporated as item 6.
In computing the overall trust measure, responses to the attention check (item 6) were discarded and the scores for the negatively phrased (reversed) items were inverted.
A participant expressing strong trust in the (X)AI assistant would thereby achieve a score of 5 on each item, yielding a cumulative trust score of 45.
Table~\ref{tab:trust-scale} summarizes the items as presented in the post-study assessment.

\begin{table}[t!]
%\scriptsize
\centering
\caption{Materials used in the post-study assessment phase. Items 1-10 show the adapted version of the rust scale adapted from \cite{hoffman2023measures}. Items marked as \textit{(Reversed)} have inverted phrasing compared to the original scale to detect contradiction in case of uniformly positive or negative response patterns. Items 5 and 9 were included as a quality control measure to detect contradictory responses.
The last two items show the optional open-ended questions used for qualitative analysis.}
\label{tab:trust-scale}
\begin{tabularx}{\textwidth}{P{0.12\textwidth}p{0.7\textwidth}P{0.17\textwidth}}
\toprule
Item No & Item & Item No in \cite{hoffman2023measures} \\ 
\midrule
1 & I am confident in the AI Assistant. I feel that it works well. & 1\\
2 & The outputs of the AI Assistant are not at all predictable. \textit{(Reversed)} & 2 \\
3 & The AI Assistant is very reliable. I can count on it to be correct all the time. & 3\\
4 & I feel safe that when I rely on the AI Assistant I will make the right decision. & 4\\
5 & The AI Assistant can perform the task worse than a novice human user. \textit{(Reversed; repeated with reversed wording to assess data quality.)} & 7\\
6 & Please select “I disagree strongly” to confirm you're paying attention. \textit{(Attention item)} & ..\\
7 & I am wary of the AI Assistant. & 6\\
8 & The AI Assistant is efficient in that it works very quickly. & 5\\
9 & The AI Assistant can perform the task better than a novice human user. \textit{(Repeated with reversed wording to assess data quality.)} & 7\\
10 & I do not like using the AI Assistant for decision making. \textit{(Reversed)} & 8\\[1ex]
\hline
& Did you notice anything about the AI tool? (optional) \textit{(Open-ended)} & ..\\[1ex]
& What do you think we wanted to find out here? (optional) \textit{(Opend-ended)} & ..\\[1ex]
\bottomrule
\end{tabularx}
\end{table}

\subsection{Participant Screening and Recruitment}
Participants were recruited via the platform Prolific\footnote{\url{https://www.prolific.com}} using a staged, sequential design targeting 4 distinct groups of 90 participants per condition, explicitly excluding those who had already participated in earlier runs.
This approach ensured independent samples for each experimental condition. 
As our study used AI-generated headshots representing three ethnic groups (i.e., job candidates of European, African, or East Asian descent), we prescreened participants to achieve a balanced ethnic representation such that each condition sample included approximately 30\% Black, 30\% White, and 30\% East Asian participants to mitigate potential confounds from ethnic bias.
Additionally, we applied another screener ensuring that each condition sample was balanced by gender (50\% male and 50\% female), which was particularly important since gender bias was one of our key experimental manipulations.
We further restricted participation to individuals fluent in English, to guarantee comprehensive understanding of the study instructions, and only allowed those with an approval rate of at least 99\% to maintain high data quality.

The target sample size of $N$=360 (90 participants per group) was determined through an \textit{a priori} power analysis using G*Power~\cite{faul2007g}.
The analysis assumed an alpha level of 0.050, power (1-$\beta$) of 0.950, four groups assessed over three measurements, with a correlation among repeated measurements of 0.500, and nonsphericity correction of 1.
For a small effect size ($f$ = 0.100; \cite{cohen1992statistical}), the analysis indicated a required sample size of 352 (i.e., 88 participants per group).
For comparison, medium effect size ($f$ = 0.250) and large effect size ($f$ = 0.400) would have required 60 and 28 participants per group, respectively. We chose to power our study for detecting small effects to ensure sufficient sensitivity to be able to identify potentially subtle effects.

\subsection{Data Quality Criteria}\label{subsec:DataQual}

To ensure the validity and reliability of our findings, we implemented a set of \textit{a priori} quality criteria.
There is an inherent risk of receiving low-quality data in online experiments due to inattentive responses, automated behavior, or technical issues.
Therefore, we established strict quality criteria from the outset to minimize measurement error and enhance the robustness of our results. 
These measures included identifying participants who completed the study too quickly (``speeders'') or took an unusually long time (``dawdlers''), defined as study times $>$ 3 standard deviations from the population mean.
We also flagged participants for ``straight-lining'' if they selected exclusively the left or right image for more than 10 consecutive trials either before or after (but not during) (X)AI interaction, or if they provided uniform, same-valence responses throughout the post-study survey phase.
Additionally, we flagged participants that failed the attention item (item 6) in the survey.
As a last quality measure, we included a repeated item with reversed wording in the survey (item 5 and item 9 in Table \ref{tab:trust-scale}) and flagged participants who responded with the same valence to both items, thus giving contradictory replies.
To maintain rigorous data quality standards, we conservatively exclude any participant flagged for even a single quality concern.

\subsection{Statistical Analysis}

We performed all statistical analyses using R (version 4.4.2)~\cite{R_core_2024}, with the experimental condition - Female Bias + AI recommendations (\textit{FB-AI}), Male Bias + AI recommendations (\textit{MB-AI}), Female Bias + XAI recommendations (\textit{FB-XAI}), and Male Bias + XAI recommendations (\textit{MB-XAI}) - serving as the independent variable.
Distributional differences in demographic covariates were evaluated using $\chi^2$ tests.

The bias shift (BS) in participant behavior was computed as: $BS = MB_{post} - MB_{pre}$, where $MB_{post}$ is the male bias in the post-(X)AI-interaction phase, and $MB_{pre}$ is the male bias in the pre-(X)AI-interaction phase.
Positive values indicate a shift toward male bias, while negative values indicate a shift toward female bias.
To evaluate the bias shift, the proportion of (X)AI alignment during the interaction phase, and the accumulated trust scores derived from the post-study assessment, we fitted separate 2×2 linear models with factors \textit{biased gender} (FB vs. MB) and \textit{XAI} (AI vs. XAI). 

For the longitudinal analysis of confidence, measured after each of the three decision phases, we employed a linear mixed-effects model using the lme4 package (v.1.1.36)\cite{bates_fitting_2015}.
This model incorporated fixed effects for group, phase, and their interaction, and included a by-subject random intercept to account for within-participant correlations~\cite{detry_analyzing_2016,muth_alternative_2016}.
Model comparisons were performed using the analysis of variance function from base R.

Last, qualitative data were described but not statistically evaluated, as they are non-quantitative in nature.

\begin{table}
  \caption{Demographic information of participants and statistical comparisons. Group differences were evaluated using $\chi^2$ tests.}
  \label{tab:Participants}
  %\resizebox{(\textwidth-\columnsep)/2}{!}{%
%\begin{tabular}{lllllll} 
\begin{tabularx}{\textwidth}{p{0.15\textwidth}p{0.15\textwidth}p{0.16\textwidth}p{0.15\textwidth}p{0.16\textwidth}p{0.09\textwidth}p{0.09\textwidth}}
\toprule
    & \textit{FB-AI} & \textit{MB-AI} & \textit{FB-XAI} & \textit{MB-XAI} & \textit{$\chi^2$} & \textit{p} value\\
\hline
\multicolumn{7}{l}{Before applying data quality criteria (\textit{N} = 363)}\\ %& & \multicolumn{4}{l}{After quality criteria (\textit{N} = 39)}\\
\cline{2-6} %\cline{6-10} 
\hline
\textit{N}   &  91 & 91 & 90 & 91 & .. & .. \\ %& & 19 & 20 & .. & .. \\
Gender$^a$ & 46f/45m & 40f/50m/1nb & 48f/42m & 45f/45m/1nb & 3.495 & .745 \\ %& & 4f/15m & 5f/15m & 182.5 & .788 \\
Age (\textit{Mdn})$^b$ & 25--34y & 25--34y & 25--34y & 25--34y & 16.911 & .324 \\ %& & 35--44y & 35--44y & 143 & .168 \\
\hline
\multicolumn{7}{l}{After applying data quality criteria (\textit{N} = 294)}\\
\hline
\textit{N}   &  73 & 69 & 70 & 82 & .. & .. \\ %& & 19 & 20 & .. & .. \\
Gender$^a$ & 34f/39m & 31f/37m/1nb & 40f/30m & 42f/40m & 5.618 & .467 \\ %& & 4f/15m & 5f/15m & 182.5 & .788 \\
Age (\textit{Mdn})$^b$ & 25--34y & 25--34y & 25--34y & 25--34y & 17.451 & .293 \\ %& & 35--44y & 35--44y & 143 & .168 \\
\hline
\multicolumn{7}{l}{Participants with balanced aptitude cases in pre- and post-phases (\textit{N} = 130)}\\
\hline
\textit{N}   &  31 & 31 & 31 & 37 & .. & .. \\ %& & 19 & 20 & .. & .. \\
Gender$^a$ & 13f/18m & 15f/15m/1nb & 18f/13m & 18f/19m & 4.863 & .561 \\ %& & 4f/15m & 5f/15m & 182.5 & .788 \\
Age (\textit{Mdn})$^b$& 25--34y & 25--34y & 25--34y & 25--34y & 17.442 & .293 \\ %& & 35--44y & 35--44y & 143 & .168 \\
\bottomrule
%\multicolumn{7}{l}{$^a$ non-parametric Wilcoxon-Mann-Whitney \textit{U} test}\\
\multicolumn{7}{l}{$^a$ f = female, m = male, nb = non-binary}\\
\multicolumn{7}{l}{$^b$ \textit{Mdn} = median age band (ranges: 18-24y, 25-34y, 35-44y, 45-54y, 55-64y,}\\
\multicolumn{7}{l}{
\hphantom{$^b$ }65y and over)}
%\end{tabular}%}
\end{tabularx}
\end{table}

\section{Results}

\subsection{Participant Flow}

We collected participant data using a staged, sequential design, targeting 4 distinct groups of 90 participants per condition.
Due to 3 participants completing the study on our server but experiencing a time-out on Prolific before marking the study as completed, we acquired data from a total of 363 participants, with $n$=91 participants in the \textit{FB-AI} condition, $n$=91 in the \textit{MB-AI} condition, $n$=90 participants in the \textit{FB-XAI} condition, and $n$=91 participants in the \textit{MB-XAI} condition.
Informed electronic consent was obtained from every participant via a clickwrap agreement before they took part in the study.
Mean study completion time was 27 minutes and 53 seconds ($\pm$12 minutes and 47 seconds \textit{SD}).
All participants received a reward of GBP 4.50 for participation.

Demographic data (age and gender) were collected and $\chi^2$ tests confirmed that the groups did not significantly differ in terms of these covariates (Table~\ref{tab:Participants}).
Quality control procedures then removed participants with problematic response patterns (0 ``speeders'', 3 ``dawdlers'', 3 failing the attention item in the survey, 9 participants straight-lining in decisions before or after AI interaction, 2 straight-liners in the survey, and 52 with contradictory survey responses), resulting in a final clean dataset of 294 participants that was used to investigate participant's alignment with the AI recommendations, reported trust and confidence in their decision-making. 
For the analysis of bias shift exhibited by participants from pre- to post-interaction phases, only decisions on cases where both candidates had an equal total aptitude score were considered, as only the bias shown in those cases can be considered to be objectively free from the confound of displayed aptitude.
Thus, the analysis of bias shift was based on a subsample of 130 participants (FB-AI: $n$=31; MB-AI: $n$=31; FB-XAI: $n$=31; MB-XAI: $n$=31;). 
Table \ref{tab:Participants} summarizes all demographic data and respective statistical comparisons of the different cohorts. In the interest of reproducible research, study data is available at: \url{https://github.com/ukuhl/BiasBackfiresXAI2025}

\subsection{Alignment with (X)AI Recommendations}
\begin{figure}[t]
\centering
\includegraphics[width=0.75\textwidth]{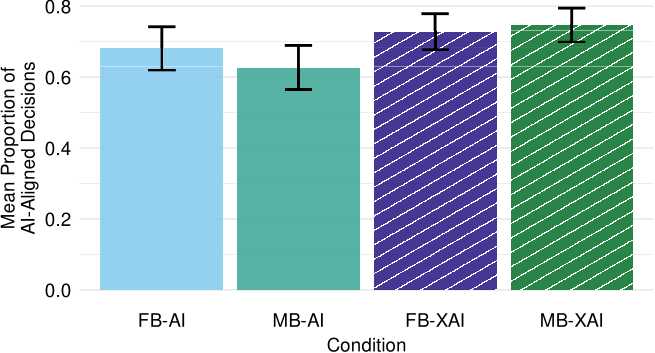}
\caption{Mean proportion of (X)AI-aligned decisions in intervention phase, stratified by condition. Whiskers represent the standard error of the mean.} \label{fig:H1_AI_alignment}
\end{figure}

% Here goes H1:
Regarding Hypothesis 1, mean proportion of alignment varied between 63 and 75\% (Fig. \ref{fig:H1_AI_alignment}).
We conducted a 2\texttimes2 ANOVA on the proportion of (X)AI alignment with factors \textit{biased gender} and \textit{XAI}.
Neither the main effect of \textit{biased gender} (\textit{F}(1,202) = 0.022, \textit{p} = 0.882) nor \textit{XAI} (\textit{F}(1,202) = 2.393, \textit{p} = .123) reached significance.
The interaction between factors \textit{biased gender} and \textit{XAI} was also non-significant (\textit{F}(1,202) = 0.435, \textit{p} = .510).
These results do not support the hypothesis that XAI CEs increase alignment with XAI recommendations compared to black-box AI recommendations.

\subsection{Bias Shift}
% Here goes H2:
%Interaction with biased AI recommendations will shift participants' gender-based decision patterns in subsequent independent evaluations, increasing alignment with the AI's bias direction.

\begin{figure}[ht]
\includegraphics[width=\textwidth]{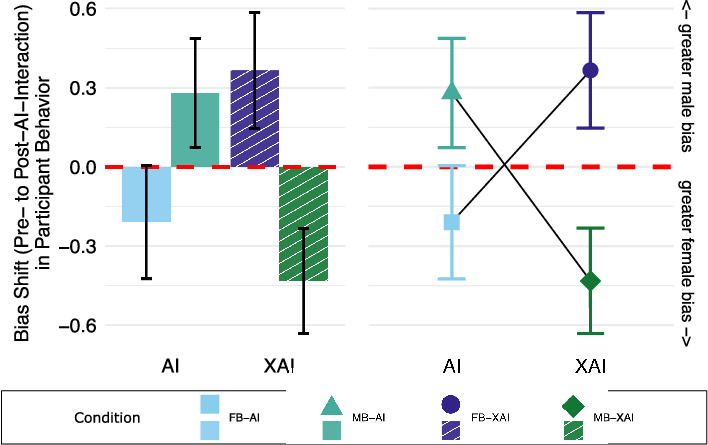}
\caption{Mean bias shift in participant behavior from pre- to post-(X)AI interaction, stratified by condition. Left: Bar plot depicting the mean bias shift for each group. Right: Interaction plot illustrating the significant interaction between factors \textit{biased gender} and \textit{XAI}. The dashed red line denotes the null line (i.e., no bias shift); values above indicate a shift toward higher male bias, while values below indicate a shift toward higher female bias. Whiskers represent the standard error of the mean.} \label{fig:H2_BiasShift}
\end{figure}

We assumed that exposure to biased (X)AI recommendations will change in participant's inherent decision bias after interaction (Hypothesis 2).
To evaluate this hypothesis, we conducted an analysis of the bias shift shown by participants from pre- to post-(X)AI-interaction phases.
While this analysis showed no significant main effect of \textit{biased gender} (\textit{F}(1, 126) = 0.782, \textit{p} = .378), nor a main effect of \textit{XAI} (\textit{F}(1, 126) = 0.206, \textit{p} = .650), it revealed a significant interaction between both factors (\textit{F}(1, 126) = 9.343, \textit{p} = .003). 

Strikingly, participants who did not receive explanations shifted their inherent biases in the same direction as the bias displayed by the AI, whereas those who received CEs shifted their bias in the opposite direction of the AI bias:
Fig. \ref{fig:H2_BiasShift} shows that in the non-explanation (AI) condition, participants exposed to female-biased AI (FB-AI, light blue square) showed a negative bias shift (toward more female-favoring decisions), while those exposed to male-biased AI (MB-AI, green triangle) showed a positive bias shift (toward more male-favoring decisions). This indicates that, without explanations, participants adopted bias patterns that aligned with the AI system's bias direction.

In contrast, when CEs were provided (XAI condition), the opposite pattern emerged. Participants exposed to female-biased XAI (FB-XAI, dark blue circle) shifted toward more male-favoring decisions, while those exposed to male-biased XAI (MB-XAI, dark green diamond) shifted toward more female-favoring decisions.

This significant interaction effect suggests that CEs play a crucial modulatory role in how algorithmic bias influences human decision-making.
Rather than simply preventing bias adoption, explanations appear to trigger a reversal effect, causing participants to shift their decision patterns in the direction opposite to the AI's bias. This finding has important implications for how XAI systems might influence human judgment in high-stakes decision contexts.

\subsection{XAI Impact on Confidence}
% Here goes H3:
%Decision confidence will vary across experimental phases, depending on the experimental manipulation.

To test the hypothesis that decision confidence varies across experimental phases (pre-AI, with AI, post-AI) and that this pattern is modulated by the experimental manipulations (\textit{gender bias} and \textit{XAI} conditions), we fitted a linear mixed-effects model.
This model included one within-subject factor (\textit{phase}) and the two between-subject factors \textit{XAI} and \textit{biased gender}, with a random intercept for participants to account for repeated measures.

\begin{figure}
\includegraphics[width=\textwidth]{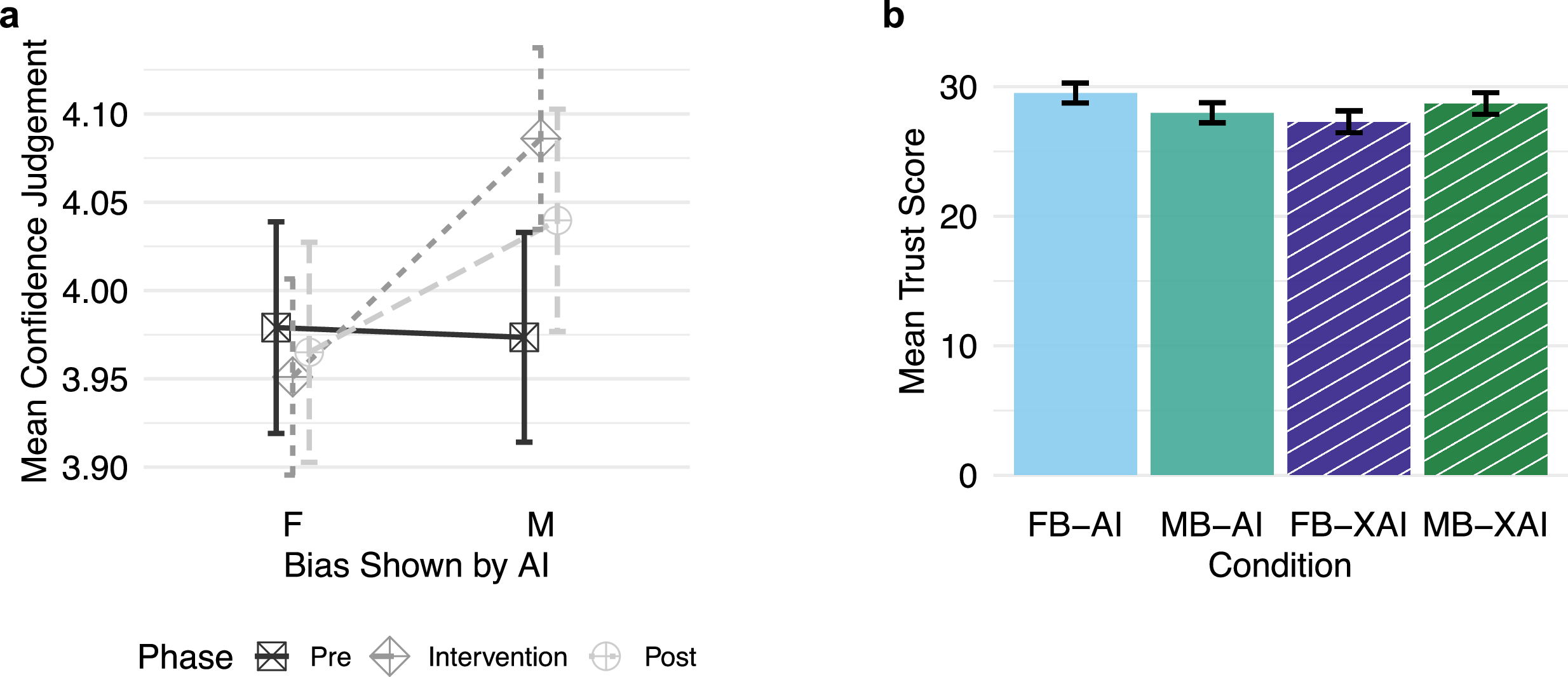}
\caption{Results on subjective measures assessed. a) Confidence: interacton between phase and gender bias. b) Trust: Mean trust score derived from post-study assessment, stratified by condition. Maximal possible value is 45. Whiskers represent the standard error of the mean.} \label{fig:H3_H4_Conf_Interaction_Trust}
\end{figure}

The analysis revealed no significant main effects of \textit{biased gender}, \textit{F}(1, 290) = 0.067, \textit{p} = .796, \textit{XAI}, \textit{F}(1, 290) = 2.587, \textit{p} = .109, or \textit{phase}, \textit{F}(2, 580) = 1.739, \textit{p} = .177.
The interaction between \textit{biased gender} and \textit{XAI} was also non-significant, \textit{F}(1, 290) = 1.394, \textit{p} = .239, as was the \textit{XAI} by \textit{phase} interaction, \textit{F}(2, 580) = 0.862, \textit{p} = 0.423, and the three-way interaction, \textit{F}(2, 580) = 2.001, \textit{p} = .136.
However, the interaction between \textit{biased gender} and \textit{phase} was significant, \textit{F}(2, 580) = 4.244, \textit{p} = .015, indicating that the variation in decision confidence across phases differed depending on the induced gender bias condition (Fig. \ref{fig:H3_H4_Conf_Interaction_Trust}a).

These results suggest that while overall confidence did not differ significantly as a function of XAI or phase alone, the impact of phase on confidence was contingent upon the gender bias condition, providing partial support for our hypothesis.

%\begin{figure}
%\includegraphics[width=0.75\textwidth]{H3_userConf_correlation_conf_alignment.png}
%\caption{Confidence effect; correlation with AI alignment} \label{fig:H3_Conf_Alignment_Correlation}
%\end{figure}

\subsection{XAI Impact on Trust}
% Here goes H4:
% Participants receiving XAI counterfactual explanations will show higher rates of trust in the AI recommendations compared to those receiving black-box AI recommendations.

%\begin{figure}
%\centering
%\includegraphics[width=0.75\textwidth]%{Figures/H4_userTrust_by_Cond_stats.png}
%\caption{Mean trust score derived from post-study assessment, stratified by condition. Maximal possible value is 45. Whiskers represent the standard error of the mean.} \label{fig:H4_Trust}
%\end{figure}

To test the hypothesis that participants receiving XAI CEs would exhibit greater trust in AI recommendations compared to those receiving black-box recommendations, we conducted an analysis of variance on the summed trust scores (Fig. \ref{fig:H3_H4_Conf_Interaction_Trust}b).
The analysis revealed no significant main effect of the \textit{XAI} condition (\textit{F}(1, 290) = 0.768, \textit{p} = .382), nor of the \textit{biased gender} condition (\textit{F}(1, 290) = 0.003, \textit{p} = .959).
The interaction between \textit{biased gender} and \textit{XAI} approached significance (\textit{F}(1, 290) = 3.253, \textit{p} = .072), yet did not reach the conventional threshold for statistical significance.
Overall, these findings offer limited support for the hypothesis that XAI CEs do not significantly enhance trust in AI recommendations compared to black-box approaches.

\subsection{Qualitative Data}
% Here goes everything we observed in the open-ended questions

To complement our quantitative findings, we analyzed responses to two open-ended questions that probed participants' observations about the (X)AI tool and their hypotheses about the study's objectives. Two researchers independently coded the responses using an inductively developed coding scheme with eight categories, achieving an inter-rater reliability of 90.3\%. %.

\subsubsection{Observations About the (X)AI Tool}
65\% of the participants provided substantive observations about the (X)AI system, while the others explicitly did not notice anything or did not provide any response. Among those who shared observations, only eight participants detected any form of systematic bias in the (X)AI's recommendations. 

The most common category of comments (77 participants) concerned general assessments of the (X)AI system's performance, with participants characterizing it as ``good'', ``bad'', ``confusing'', or ``illogical'' compared to human decision-making.
51 participants attempted to reverse-engineer the (X)AI's decision-making process, hypothesizing various patterns such as preferential weighting of specific qualification criteria (e.g., education or soft skills) or positional biases (e.g., consistently favoring candidates presented on the left). 

\subsubsection{Study Purpose Recognition}
When asked about the study's objectives, participants' responses revealed varying levels of insight into its true purpose. 
Only five participants correctly identified that the study investigated how AI recommendations might influence human gender bias. An additional five participants recognized the study's focus on bias more broadly but did not specifically identify gender as the variable of interest.

A larger group of 40 participants understood that the study examined how AI influences human decision-making in general without identifying the specific focus on bias transmission. However, 75\% of the participants either misidentified the study's purpose or provided no response.

\section{Discussion}
Our findings reveal unexpectedly complex effects of AI bias and explanations on human decision-making in hiring contexts. The results raise important questions about the role of XAI in either mitigating or potentially amplifying algorithmic bias transmission. 
Extending previous work on bias internalization through AI interaction~\cite{vicente2023humans}, the current results add crucial insights about how explainability influences this process.

\subsection{Low Bias Detection Despite High AI Alignment}

Critically, only 8 out of 294 participants detected systematic bias in the (X)AI recommendations, despite every participant interacting with a demonstrably biased system.
While this may initially raise questions about participant attentiveness and data quality, research systemically examining data quality in online human-subjects assessments suggests otherwise: the robust quality controls employed by the platform used were shown to yield reliable responses~\cite{douglas2023data}, and frequently surveyed participants do not exhibit lower data quality than those surveyed less often~\cite{eisele2022effects}.
Moreover, our rigorous data quality measures (see Section~\ref{subsec:DataQual}) ensured that only responses from attentive and motivated participants were included in the final analysis.

Therefore, this finding rather serves as evidence for a ``invisible influence'' scenario that presents a significant challenge for the responsible deployment of AI in hiring contexts, confirming earlier warnings regarding the subtle manifestation of algorithmic biases~\cite{rosenthal2024michael}.
When bias operates without conscious awareness, traditional approaches to bias mitigation relying on human oversight may prove ineffective~\cite{pulivarthy2025bias}. %, a vulnerability that Pulivarthy and Whig~\cite{pulivarthy2025bias} previously identified in their analysis of seemingly neutral feature selection in hiring algorithms.

The high rate of alignment with (X)AI recommendations (70\% in cases with comparable candidates) further compounds this concern.
Participants were not simply ignoring the (X)AI recommendations, but actively incorporating them into their decision-making process, even when those recommendations contained systematic bias.
This finding extends research on automation bias, leading professionals to defer to AI suggestions even when these conflict with their own judgment~\cite{kucking2024automation}, and demonstrates transmission of algorithmic biases to human decision-makers.

\subsection{Counterfactual Explanations and Bias Reversal}

Our most striking finding concerns the differential effects of CEs on bias adoption. Without explanations, participants tended to adopt the AI bias, showing an increased preference for the AI-favored gender in subsequent independent decisions.
However, when provided with CEs, participants demonstrated a reversal effect, showing bias for the AI-disfavored gender. This finding adds critical nuance to prior work postulating that CEs enhance transparency in a manner similar to human cognition~\cite{wang2021counterfactual}, but did not anticipate this reversal phenomenon.

This unexpected pattern may suggest that CEs trigger a form of psychological reactance, where awareness of the decision-making process leads to conscious or unconscious rejection of the AI's preferences.
This may indicate an implicit awareness of the presented bias facilitated through XAI, leading to an over-compensatory adjustment in the opposite direction.

While this might initially seem positive from a bias-prevention perspective, it raises concerns about whether such reactance truly promotes more objective decision-making or simply replaces one form of bias with another. This finding challenges assumptions regarding the universally positive effects of CEs on human-AI collaborative decision-making~\cite{lee2023understanding}, suggesting a more complex relationship than previously theorized.

\subsection{Trust and Confidence Patterns}

Trust levels did not differ significantly across conditions, suggesting that trust in AI may be more stable than previously assumed, even in the face of AI bias.
Our results support similar observations regarding the complex relationship between explanations and trust~\cite{wang2021explanations,scharowski2023exploring}. This stability of trust across conditions, combined with the low rate of bias detection, indicates that users may develop and maintain trust in biased AI systems, challenging assumptions about how explainability positively calibrates trust~\cite{Biloborodova.2023}.

The variation in confidence levels observed deserves particular attention: Confidence varied as a function of study phase, but only among participants exposed to male-biased AI. 
This gender-specific effect may suggest that existing societal biases and expectations modulate the interaction between AI bias and human decision-making, potentially pointing to false confidence in AI recommendations that align with personal biases~\cite{Humer.2024}.
The finding that confidence patterns varied systematically in response to male-biased but not female-biased AI recommendations points to possible interactions with broader gender-related cognitive schemas and social norms, echoing concerns that explanations might legitimize rather than expose algorithmic bias~\cite{RasiklalYadav.2024}.

\subsection{Implications for XAI Design}

These findings have significant implications for the design and deployment of XAI systems in hiring contexts. First, they challenge the assumption that the mere presence of explainability necessarily leads to better human oversight or more objective decision-making. While explanations did prevent direct bias adoption, they appeared to promote opposition rather than objectivity. This echoes concerns that the relationship between explanation and trust requires careful calibration to avoid creating false confidence in flawed systems~\cite{Thalpage.2023}.

Second, the low rate of bias detection, despite the presence of explanations, suggests that current approaches to XAI may not be sufficient for helping users identify systematic patterns of bias. This points to the need for new explanation formats or supplementary tools specifically designed to make patterns of bias more salient to users, as suggested by Militello et al.~\cite{Militello.2025} in their work on how interface design choices significantly influence user behavior and decision-making processes. 
Our findings support Wanner et al.'s~\cite{wanner2022effect} observation that transparency can significantly influence users' willingness to rely on AI insights for their decision-making, though not always in ways that lead to objectively better outcomes.

\subsection{Qualitative Insights}

Our qualitative findings complement the quantitative results by revealing participants' limited awareness of the (X)AI's systematic biases. When asked about the (X)AI tool, most participants focused on general performance assessments or attempted to identify decision patterns based on qualification criteria rather than recognizing gender bias. This lack of bias detection mirrors findings by Kordzadeh et al.~\cite{kordzadeh2022algorithmic} regarding the challenges of detecting and mitigating algorithmic bias in everyday applications. Coupled with our finding that only five participants correctly identified the study's focus on bias transmission, this underscores how algorithmic bias can influence decision-making without conscious awareness.

The disparity between participants who detected bias (8) and those who correctly identified the study's purpose (5) suggests a disconnection between recognizing bias and understanding its potential influence on one's own decision-making. This parallels observations by Hemmer et al.~\cite{Hemmer.2024} regarding how information and capability asymmetries between humans and AI systems can affect decision quality. This is particularly concerning for real-world applications, as it indicates that users may be vulnerable to adopting algorithmic biases even when interacting with supposedly transparent AI systems, supporting concerns raised by Schemmer et al.~\cite{schemmer2023towards} about how users develop mental models of AI capabilities through experience.

\subsection{Limitations and Future Directions}

Several limitations of our study suggest directions for future research.
First, our experiment focused on gender bias in a controlled setting with artificial aptitude scores. 
We acknowledge that this study setting only represents a proxy to investigate (X)AI's influence on human decision-making with limited realism.
Future work should examine whether similar patterns emerge with other types of bias and in more naturalistic decision-making contexts, expanding on approaches suggested by Huang and Zaslavsky~\cite{Huang.2024} for detecting bias using contextual knowledge graphs.
Additionally, while we observed clear effects of CEs on bias adoption, we did not explore the specific mechanisms through which these explanations influenced decision-making.
Understanding these mechanisms could help inform more effective approaches to explanation design, potentially building on Del Ser et al.'s~\cite{DelSer.2024} work on how CEs help users develop nuanced understanding of AI reliability.

The observed reversal effect in the XAI condition particularly warrants further investigation.
Future studies might explore whether this effect persists over longer time periods and whether it generalizes to other types of decisions and biases.
In upcoming follow-up studies, we will examine how participants react to more obvious as well as more subtle biases, and whether the reversal effects observed in the current study generalize to these settings.
Additionally, research could examine whether alternative explanation formats might better support truly objective decision-making rather than simply promoting bias reversal, perhaps utilizing interactive systems like FACET~\cite{VanNostrand.2024} or CL-XAI~\cite{suffian2023cl}.
Such interactive systems may also serve to explore the potential of personalized explanations as a strategy to mitigate unwanted effects, such as participants overlooking biases.
By tailoring explanations to better align with an individual's mental model, it may be possible to relieve cognitive load and make biases more explicit, thereby encouraging more critical evaluation of AI outputs.

Finally, the gender-specific effects on confidence suggest the need for more detailed investigation of how AI bias interacts with existing social biases and expectations. This might include examining how different user demographics respond to various forms of AI bias and exploring how individual differences in attitudes toward gender and technology moderate these effects. This future direction aligns with Papenmeier's findings~\cite{papenmeier2022s} that the effect of explanations on trust is contingent upon model accuracy, suggesting a need for more nuanced understanding of these complex relationships.

\section{Conclusion}
This study provides important insights into how biased AI recommendations influence human decision-making and the complex role of CEs in this process. Our findings reveal that while participants readily incorporated (X)AI recommendations into their decisions, they rarely detected the underlying systematic bias. More critically, we discovered that CEs, while preventing direct bias adoption, led to an unexpected reversal effect where participants demonstrated bias in the opposite direction of the AI's preferences.

The interplay between algorithmic bias and explanation is nuanced: without explanations, users adopted the bias direction of the AI system in their subsequent independent decisions. When provided with CEs, however, their decisions reflected an opposite bias pattern. This finding suggests that current XAI approaches may inadvertently create reactance bordering on over-compensation, rather than promoting truly objective decision-making. The fact that overall trust did not differ significantly across conditions, despite these varying effects on decision patterns, further highlights the complex relationship between explanations, trust, and behavior.

Our observation of gender-specific confidence effects (specifically in the male-biased AI condition) points to how societal biases may interact with (X)AI biases to create differential impacts on decision certainty. Additionally, our qualitative findings revealed that most participants remained unaware of both the (X)AI's bias and its influence on their own judgment, creating an ``invisible influence'' scenario that poses significant challenges for responsible AI deployment.

These results might have substantial implications for the design and deployment of AI systems in hiring contexts and the broader field of explainable AI.
They challenge the assumption that the mere presence of explainability necessarily leads to more objective human decision-making.
Instead, our findings suggest that current approaches to XAI might need to be reconsidered to better support bias detection and prevent unwanted (over)-compensation effects on human decision-making.

Looking ahead, these insights point to the need for more nuanced approaches and more research to human-AI collaboration in high-stakes decisions.
Future work should focus on developing explanation methods that not only make AI decision processes transparent but also effectively highlight systematic patterns of bias.
Additionally, organizations implementing AI decision support systems should consider incorporating specific safeguards against both direct bias adoption and reactionary bias reversal.
As AI systems become increasingly integrated into hiring processes, understanding and managing their influence on human decision-making becomes crucial. Our research suggests that achieving this goal requires careful attention not just to the technical aspects of AI systems, but also to the complex psychological dynamics of human-AI interaction. 
The challenge of algorithmic bias is not simply a technical matter of fixing biased algorithms but also a psychological one of understanding how humans respond to, interpret, and potentially internalize or react against biased AI recommendations.

\begin{credits}
\subsubsection{\ackname}
This research was supported by the research training group Dataninja (Trustworthy AI for Seamless Problem Solving: Next Generation Intelligence Joins Robust Data Analysis) funded by the German federal state of North Rhine-Westphalia, and project KI-Akademie OWL, financed by the Federal Ministry of Education and Research Germany (BMBF) and supported by the Project Management Agency of the German Aerospace Centre (DLR) under grant no. 01IS24057A.

\subsubsection{\discintname}
The authors have no competing interests to declare that are relevant to the content of this article.
%It is now necessary to declare any competing interests or to specifically
%state that the authors have no competing interests. Please place the
%statement with a bold run-in heading in small font size beneath the
%(optional) acknowledgments\footnote{If EquinOCS, our proceedings submission
%system, is used, then the disclaimer can be provided directly in the system.},
%for example: The authors have no competing interests to declare that are
%relevant to the content of this article. Or: Author A has received research
%grants from Company W. Author B has received a speaker honorarium from
%Company X and owns stock in Company Y. Author C is a member of committee Z.
\end{credits}

\bibliographystyle{splncs04}
\bibliography{references}
%\bibitem{Rosenthal and Sach 2024}
%Rosenthal-von der Pütten, A.M., Sach, A.: Michael is better than Mehmet: exploring the perils of algorithmic biases and selective adherence to advice from automated decision support systems in hiring. Frontiers in Psychology, vol. 15, 1416504 (2024). \doi{10.3389/fpsyg.2024.1416504  }
%
%\bibitem{Pulivarthy and Whig 2024} 
%Pulivarthy, P., Whig, P.: Bias and Fairness Addressing Discrimination in AI Systems. Advances in human and social aspects of technology book series, vol. , 103–126 (2024). \doi{10.4018/979-8-3693-4147-6             } 
%Author, F.: Article title. Journal \textbf{2}(5), 99--110 (2016)
%Author, F., Author, S.: Title of a proceedings paper. In: Editor,
%F., Editor, S. (eds.) CONFERENCE 2016, LNCS, vol. 9999, pp. 1--13.
%Springer, Heidelberg (2016). \doi{10.10007/1234567890          %                    }
%Author, F., Author, S., Author, T.: Book title. 2nd edn. Publisher,
%Author, A.-B.: Contribution title. In: 9th International Proceedings
%on Proceedings, pp. 1--2. Publisher, Location (2010)
%LNCS Homepage, \url{http://www.springer.com/lncs}, last accessed 2023/10/25
%\end{thebibliography}
\end{document}